# Timing Jitter Induced by Stochastic Baseline Fluctuations in High-Count-Rate Superconducting Nanowire Single-Photon Detectors

Dianpeng Wang,[1,2,3] You Xiao,[1,2] Jiamin Xiong,[1,2] Chenrui Wang,[1,2,3] Zhen Wan,[1,2] Hongxin Xu,[1,2,3] Chaomeng Ding,[1,2,3] Jia Huang,[1,2] Lixing You,[1,2,3] Hao Li,[1,2,4,] *

[1]*Shanghai Key Laboratory of Superconductor Integrated Circuit Technology, Shanghai Institute of Microsystem and Information Technology, Chinese Academy of Sciences, Shanghai 200050, China*
[2]*National Key Laboratory of Materials for Integrated Circuits, Shanghai Institute of Microsystem and Information Technology, Chinese Academy of Sciences, Shanghai 200050, China*
[3]*Center of Materials Science and Optoelectronics Engineering, University of Chinese Academy of Sciences, Beijing 100049, China*
[4]*Shanghai Research Center for Quantum Sciences, Shanghai 201315, China*
**lihao@mail.sim.ac.cn*

**ABSTRACT**: Superconducting nanowire single-photon detectors (SNSPDs) have demonstrated timing jitter in the few-picosecond regime, yet their timing resolution deteriorates substantially under high-count-rate operation. Existing interpretations mainly attribute this degradation to deterministic waveform distortions, such as multiphoton responses and pulse pile-up, yet the experimentally observed jitter broadening at high count rates cannot be fully accounted for within this picture. Here, we show that stochastic baseline fluctuations arising from finite-memory readout dynamics constitute an intrinsic source of the count-rate-dependent timing jitter in SNSPD systems. For stochastically arriving photons, overlapping recovery responses accumulate in the readout chain and generate statistically fluctuating baselines, which are converted into timing uncertainty through threshold-based timing extraction. We develop a stochastic-process framework that quantitatively connects photon statistics, readout dynamics, and timing jitter. The framework predicts characteristic scaling behaviors, including a nonmonotonic dependence of baseline fluctuations under pulsed excitation with a maximum near half of the repetition frequency. These predictions are quantitatively verified through systematic variations of count rate, circuit time constant, and detector dynamical properties. Our results identify stochastic baseline dynamics as a fundamental mechanism limiting timing resolution in high-count-rate SNSPD operation and provide a general framework for optimizing finite-memory high-speed photon-counting systems.

## I. INTRODUCTION

Timing jitter is a key performance metric of single-photon detectors which quantifies the temporal uncertainty in the photon arrival time recorded by the detector. Superconducting nanowire single-photon detectors (SNSPDs)[1] have achieved timing jitter in the few-picosecond regime[2-7], establishing them as one of the most precise photon-timing technologies currently available. Continuous advances in nanowire engineering, device fabrication, and cryogenic readout techniques have progressively reduced the timing jitter of SNSPD systems. As the achievable timing resolution approaches the few-picosecond level, increasing attention has shifted toward system-level mechanisms that govern timing performance under practical operating conditions. In particular, maintaining low timing jitter at high count rates (tens of Mcps and above) has become a central challenge for high-speed photon-counting applications, including quantum information processing[8-17], deep-space optical communication[18-20], and single-photon ranging[5, 21, 22].

Experimentally, timing jitter is well-known to increase substantially as the count rate rises [20, 23-25]. Existing interpretations mainly attribute this degradation to deterministic waveform distortions, including multiphoton responses and pulse pile-up effects[23, 26-28], which modify the pulse shape and alter threshold-crossing dynamics. These mechanisms have provided important insight into high-count-rate timing behavior. However, experimental observations consistently show that substantial timing broadening persists even when such deterministic effects are strongly suppressed, suggesting that an additional physical mechanism contributes to the observed degradation.

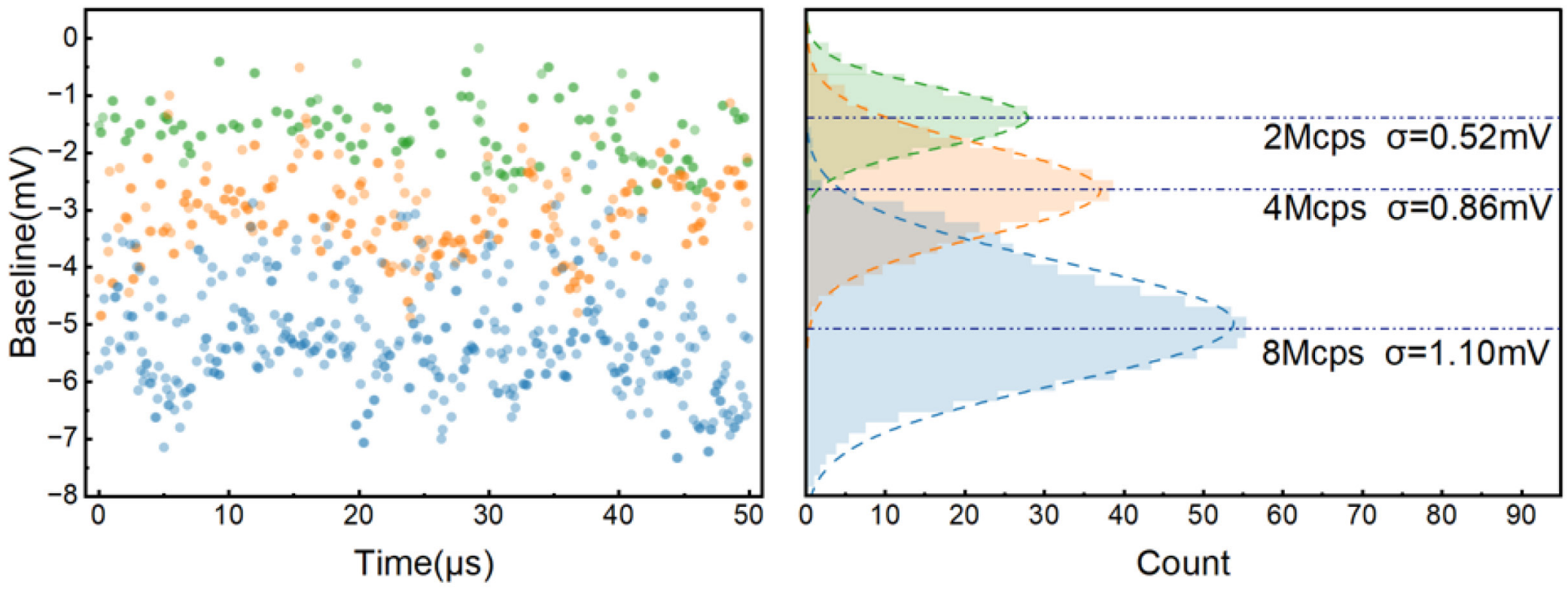

FIG. 1. Experimental evidence of stochastic baseline fluctuations induced by finite-memory readout dynamics in SNSPD systems. Left panel: instantaneous pre-event baseline values measured over a 50 μs acquisition window at different count rates, where each point corresponds to the baseline level immediately preceding a detection event. Right panel: corresponding statistical distributions obtained from the same baseline data. As the count rate increases, the baseline exhibits progressively stronger fluctuations accompanied by systematic broadening of the corresponding distributions. These observations directly reveal the stochastic accumulation of overlapping finite-memory responses in the readout chain under high-count-rate operation.

At the same time, the low-jitter SNSPD systems are commonly implemented using AC-coupled or effectively high-pass-filtered readout architectures to suppress low-frequency noise and stabilize broadband amplification[2-7]. Such readout schemes inherently possess finite-memory dynamics: each detection event produces a recovery response that persists over a finite temporal window. Under stochastic photon arrival, these residual responses accumulate statistically and continuously perturb the signal baseline. As the count rate increases, the baseline therefore evolves from a static reference level into a stochastic dynamical variable governed by the collective history of prior detection events.

Figure 1 experimentally illustrates this behavior. As the count rate increases, the pre-event baseline exhibits progressively stronger stochastic fluctuations, accompanied by a systematic broadening of the corresponding statistical distributions. These observations directly reveal the growth of baseline variance arising from the accumulation of overlapping finite-memory responses. Because threshold-based timing extraction depends sensitively on the instantaneous baseline condition, such fluctuations are directly converted into timing uncertainty. Under high-count-rate operation, the resulting jitter broadening can become comparable to or even exceed the low-count-rate system timing jitter.

In this work, we identify stochastic baseline fluctuations as an intrinsic mechanism governing timing-jitter broadening in high-count-rate SNSPD systems. We show that finite-memory dynamics in the readout chain transform stochastic photon arrivals into fluctuating baseline trajectories, which are subsequently converted into timing uncertainty through threshold-based timing extraction. Based on this physical picture, a stochastic theoretical framework is developed to describe the interplay among photon statistics, system memory, and timing performance under both continuous-wave and pulsed excitation. Systematic experiments performed across different count rates, recovery timescales, and detector dynamical regimes confirm the predicted scaling behavior and clarify the conditions under which stochastic baseline dynamics dominate the timing performance. These results establish a general physical framework for understanding timing limitations in high-speed photon-counting systems.

## II. PHYSICAL PICTURE

The physical origin of the observed stochastic baseline fluctuations can be understood from the finite-memory response characteristics of the SNSPD readout chain, as illustrated in Fig. 2. In practical SNSPD systems, AC coupling or effectively high-pass-filtered amplification is widely employed to suppress low-frequency noise and stabilize broadband readout operation. Such readout architectures impose a charge-balancing constraint on the electrical response, such that each detection pulse is accompanied by a compensating baseline relaxation[29], as schematically illustrated in Fig. 2(a). Consequently, a detection event cannot produce a strictly localized temporal response, but instead generates a residual perturbation that persists over a finite recovery time constant. Following a single detection event, the baseline relaxes gradually toward equilibrium over a characteristic time constant determined by the effective recovery dynamics of the readout system, as illustrated in Fig. 2(b). This finite relaxation process introduces temporal memory into the system, causing the instantaneous baseline to depend on the cumulative history of preceding

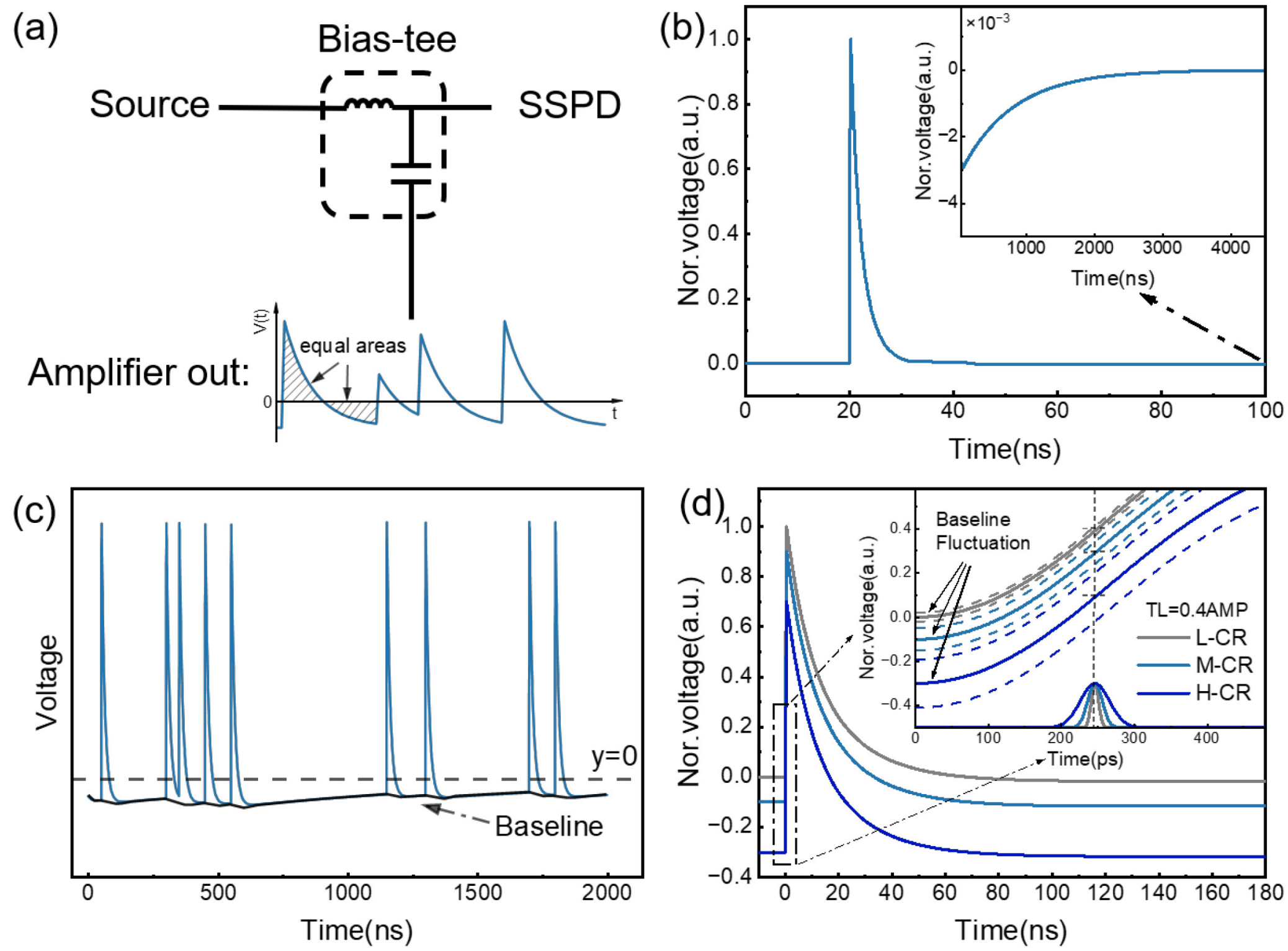

FIG. 2. Physical picture of stochastic baseline fluctuations and timing jitter broadening in finite-memory SNSPD readout systems. (a) Schematic illustration of the finite-memory response introduced by an AC-coupled or effectively high-pass-filtered SNSPD readout chain. The bias-tee is typically composed of a broadband inductor in the DC bias arm and a coupling capacitor in the RF output arm. Together with the input capacitance and input impedance of the subsequent broadband amplifier, the coupling capacitor determines the low-frequency cutoff and effective recovery timescale of the readout system. Each detection event therefore generates a compensating baseline displacement that relaxes over a finite temporal window. (b) Temporal evolution of the baseline following a single detection event. Owing to the finite recovery time constant $\tau_e$, the baseline does not return instantaneously to equilibrium, but instead exhibits a gradual recovery governed by the readout dynamics. The inset shows the long-timescale recovery process following the SNSPD rising edge, illustrating the slowly relaxing baseline tail induced by the finite-memory response. (c) Under finite count-rate operation, residual baseline perturbations from successive detection events overlap and accumulate statistically, driving the baseline into a fluctuating stochastic state. (d) Conversion of stochastic baseline fluctuations into timing jitter in threshold-based timing extraction. Variations in the instantaneous baseline modify the effective threshold-crossing condition and introduce additional timing uncertainty. The inset illustrates representative pulse waveforms under low-count-rate (L-CR), medium-count-rate (M-CR), and high-count-rate (H-CR) operation, showing progressively stronger baseline fluctuations that shift the apparent pulse onset and broaden the measured timing distribution under fixed-threshold discrimination.

detection events rather than solely on the current pulse.

When the detector operates at a finite count rate, successive detection events occur within this recovery window. As illustrated schematically in Fig. 2(c), the residual responses associated with different events therefore accumulate and overlap in time. Under stochastic photon arrival, this accumulation process drives the baseline into a fluctuating dynamical state whose statistical properties are governed jointly by the photon-arrival statistics and the finite-memory response of the readout chain. The baseline thus evolves from a static reference level into a stochastic variable with history-dependent fluctuations.

These stochastic baseline fluctuations directly affect timing extraction in threshold-discrimination-based SNSPD systems. Because the photon arrival time is determined by a fixed voltage threshold, fluctuations in the instantaneous baseline modify the effective threshold-crossing condition, thereby converting baseline-voltage fluctuations into additional timing uncertainty, as illustrated in Fig. 2(d). Unlike conventional multiphoton-response or pulse-pile-up mechanisms, which arise primarily from deterministic waveform distortions associated with individual pulse interactions, the present mechanism originates from the collective statistical accumulation of finite-memory responses under stochastic excitation.

This physical picture establishes the basis for a stochastic-process description of baseline dynamics in finite-memory SNSPD systems. In the following section, we develop a quantitative framework that connects photon-arrival statistics, readout response dynamics, and timing jitter broadening under both continuous-wave and pulsed excitation conditions.

## III. THEORY

To quantitatively describe stochastic baseline fluctuations in high-count-rate SNSPD systems, the readout dynamics are treated as a finite-memory linear system driven by stochastic photon arrivals. Under strongly attenuated optical excitation, photon detection events are well described by a Poisson point process[30, 31] with average count rate $\lambda$. In the experimentally relevant regime considered here, detection events are statistically independent and the baseline dynamics are dominated by the linear superposition of residual recovery responses from preceding events.

Each detection event generates a finite-memory response in the readout baseline. The baseline dynamics can therefore be described as a stochastic shot-noise process formed by the superposition of all preceding responses. The single-event response kernel is described by

$$h(t) = \frac{Q}{\tau_e} \cdot e^{-\frac{t}{\tau_e}} \cdot u(t) \tag{1}$$

Where $Q$ is the effective pulse area associated with a single detection event, $\tau_e$ is the recovery time constant of the readout system, and $u(t)$ is the Heaviside step function.

For a sequence of photon arrival times $t_i$, the instantaneous baseline is given by

$$B(t) = \sum_i h(t - t_i) \tag{2}$$

which describes a Poisson-driven finite-memory stochastic process.

Under Poisson excitation, the stationary baseline statistics can be derived analytically using standard shot-noise theory. The ensemble-averaged baseline is

$$E\big(B(t)\big) = \int_{-\infty}^{t} \lambda \cdot h(t - t')dt' = \lambda Q \tag{3}$$

while the baseline variance follows directly from Campbell's theorem[32]:

$$Var\big(B(t)\big) = \lambda \int_0^{+\infty} h^2(t)\, dt = \frac{\lambda Q^2}{2\tau_e} \tag{4}$$

leading to a baseline standard deviation

$$\sigma_B = Q \cdot \sqrt{\frac{\lambda}{2\tau_e}} \tag{5}$$

This scaling law indicates that baseline fluctuations increase with event rate and decrease with recovery time constant, reflecting the intrinsic memory effect of the readout circuit.

For pulsed excitation, the stochastic process becomes discretized by the laser repetition period $T = 1/f_{rep}$, In this regime, photon arrivals follow a Bernoulli process with detection probability $p = \lambda/f_{rep}$ per excitation cycle. The resulting baseline evolution obeys a discrete stochastic recurrence relation (See Appendix B), yielding the steady-state fluctuation amplitude

$$\sigma_B = \frac{Q}{\tau_e \cdot \sqrt{e^{\frac{2T}{\tau_e}} - 1}} \cdot \sqrt{\frac{\lambda}{f_{rep}}\left(1 - \frac{\lambda}{f_{rep}}\right)} \tag{6}$$

In the limit $T \ll \tau_e$, this reduces to

$$\sigma_B = \frac{Q}{\sqrt{2\tau_e}} \cdot \sqrt{\lambda\left(1 - \frac{\lambda}{f_{rep}}\right)} \tag{7}$$

while in the limit $T \to \infty$, the result recovers the continuous excitation case.

Importantly, the theory predicts a characteristic nonmonotonic dependence of baseline fluctuations under pulsed excitation. The fluctuation amplitude reaches a maximum near $\lambda \approx f_{rep}/2$, reflecting the competition between stochastic event randomness and the increasing regularity of the excitation sequence at high detection probability. In the low-count-rate regime ($\lambda \ll f_{rep}$), detection events are sparse and residual responses overlap weakly, resulting in small baseline fluctuations. As $\lambda$ increases, stochastic overlap between successive recovery responses enhances the baseline variance. In contrast, as $\lambda \to f_{rep}$, nearly every excitation pulse produces a detection event, causing the sequence to approach a quasi-deterministic periodic train and thereby suppressing stochastic fluctuations.

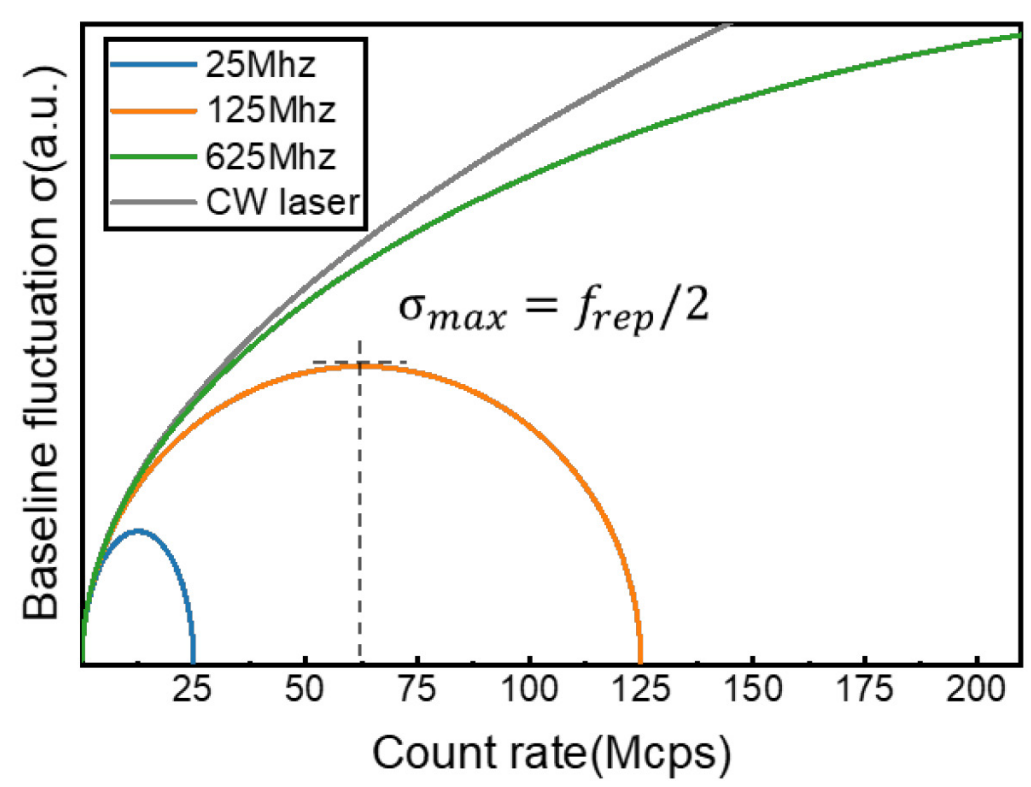

FIG. 3. Scaling behavior of stochastic baseline fluctuations under continuous-wave and pulsed excitation. Calculated standard deviation of the baseline fluctuations as a function of detection rate for continuous-wave and pulsed excitation conditions. Under continuous excitation, the baseline variance increases monotonically with count rate due to the statistical accumulation of overlapping finite-memory responses. Under pulsed excitation, the discrete repetition period modifies the fluctuation statistics and produces a nonmonotonic dependence with a maximum near one-half of the repetition frequency. As the repetition frequency increases, the pulsed-excitation results continuously approach the continuous Poisson limit, demonstrating the unified stochastic description of both operating regimes.

The theoretical predictions for continuous-wave and pulsed excitation are summarized in Fig. 3. Under continuous excitation, the baseline fluctuations increase monotonically with count rate due to the cumulative accumulation of finite-memory responses. Under pulsed excitation, however, the finite repetition constraint produces a distinct nonmonotonic dependence. As the repetition frequency increases, the pulsed-excitation results continuously converge toward the continuous Poisson limit, demonstrating that both excitation regimes are naturally unified within the same stochastic framework.

Timing extraction in SNSPD systems is typically performed using fixed-threshold discrimination. Baseline fluctuations are therefore converted into timing uncertainty through the local pulse slope $S = dV/dt$ according to

$$\sigma_t = \frac{\sigma_B}{S} \tag{8}$$

Thus, the scaling behavior of timing jitter directly inherits that of the baseline fluctuations, establishing a quantitative link between readout statistics and timing performance.

Under continuous excitation, using $Q = A\tau_s$ and $S = A/t_{rise}$, this leads to

$$\sigma_{t_{CW}} = \frac{t_{rise} \cdot \tau_s}{\sqrt{2\tau_e}} \sqrt{\lambda} \tag{9}$$

while under pulsed excitation,

$$\sigma_{t_{pulse}} = \frac{t_{rise} \cdot \tau_s}{\sqrt{2\tau_e}} \cdot \sqrt{\lambda\left(1 - \frac{\lambda}{f_{rep}}\right)} \tag{10}$$

Where $t_{rise}$ is the rise time of the SNSPD pulse, $\tau_s$ is the decay time constant of the SNSPD pulse, $\tau_e$ is the recovery time constant of the readout circuit, $A$ is the amplitude of the pulse, $\lambda$ is the count rate, and $f_{rep}$ is the repetition frequency of the pulsed laser. These expressions provide a direct theoretical prediction for the count-rate dependence of timing jitter.

## IV. EXPERIMENTAL VALIDATION

To quantitatively compare the theoretical predictions with experiment, the stochastic-process parameters are related to experimentally measurable pulse characteristics of the SNSPD readout system. For the extraction of theoretical prediction parameters, please see Appendix D.

### A. Count-rate dependence under pulsed excitation

We first experimentally examine the predicted count-rate scaling of stochastic baseline fluctuations under pulsed excitation. The detector count rate is continuously tuned by varying the incident optical power while maintaining a fixed laser repetition frequency. Output waveforms are recorded using a high-bandwidth oscilloscope, from which baseline statistics are extracted, while timing jitter is independently measured using a time-to-digital converter (TDC). The experiment is performed using an SNSPD with a recovery time constant of approximately 11.4 ns, operated at a bias current of 19 μA. The detector is driven by a pulsed laser with a repetition frequency of 20 MHz (See APPENDIX A for more information about the experimental device and APPENDIX F for the verification of 125Mhz pulsed laser and continuous-wave (CW) laser), and an external coupling capacitor of 10 nF is employed in the readout chain. The theoretical prediction shown in Fig. 4(a) is calculated directly from independently measured system parameters without adjustable fitting coefficients.

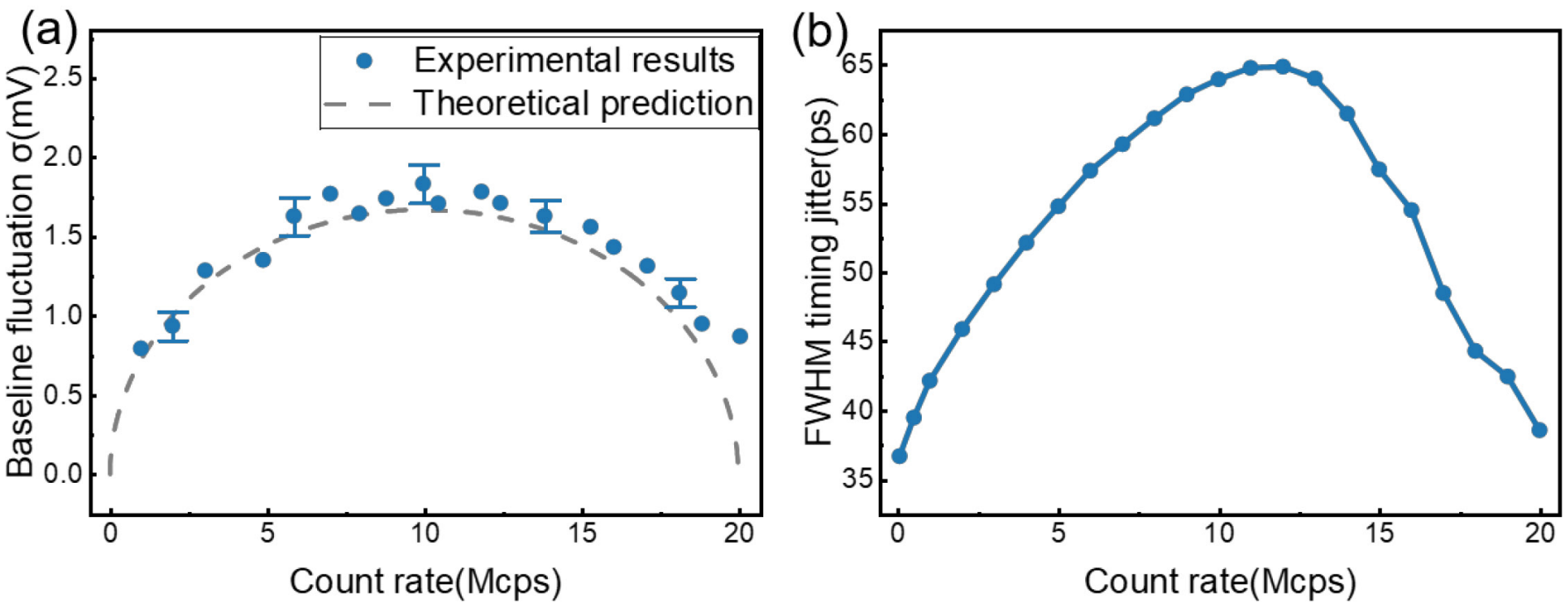

FIG. 4. Experimental validation of stochastic baseline fluctuations and timing jitter broadening under pulsed excitation. Error bars denote the standard deviation obtained from repeated measurements (N = 6). (a) Measured standard deviation of the pre-event baseline as a function of count rate under pulsed excitation. The experimental results quantitatively reproduce the predicted nonmonotonic dependence, with a maximum near one-half of the laser repetition frequency. The dashed curve represents the theoretical prediction calculated from independently measured system parameters without adjustable fitting coefficients. (b) Measured timing jitter full width at half maximum (FWHM) of the single-photon peak as a function of count rate. The single-photon contribution is extracted from the multi-peak timing histogram using the procedure described in APPENDIX D. The timing jitter exhibits the same count-rate dependence as the baseline fluctuations, demonstrating the direct conversion of stochastic baseline fluctuations into timing uncertainty.

As shown in Fig. 4(a), the measured baseline fluctuations exhibit a pronounced nonmonotonic dependence on count rate. The fluctuation amplitude initially increases with count rate, reaches a maximum near half of the repetition frequency, and decreases at higher rates. The measurements quantitatively follow the theoretical prediction for pulsed stochastic excitation, directly confirming the finite-memory accumulation mechanism predicted by the stochastic framework. The slight discrepancy that the experimental measurements are marginally higher than the theoretical predictions arises from the inherent baseline noise of the readout system, which exhibits a standard deviation of approximately 0.45 mV.

Furthermore, the timing jitter extracted from the single-photon response peak (see APPENDIX E for the extraction procedure) is shown in Fig. 4(b). The single-photon component is analyzed separately in order to exclude additional timing broadening associated with multiphoton detection events[26, 28]. The measured timing jitter follows the same count-rate dependence as the baseline fluctuations, indicating that stochastic baseline dynamics are converted into timing uncertainty through threshold-based timing extraction.

### B. Influence of circuit time constant

To isolate the role of finite-memory recovery dynamics, the effective recovery time constant $\tau_e$ is systematically tuned by varying the coupling capacitance (the actual capacitors we use are 2.2, 4.7, 10, 47 and 100 nF) in the AC-coupled readout chain. The experiment is performed using an SNSPD with a recovery time constant of approximately 11.4 ns, operated at a bias current of 19 μA. The detector is driven by a pulsed laser with a repetition frequency of 20 MHz, and the counting rate of SNSPD is fixed at 10Mcps. The theoretical prediction shown in Fig. 5(a) is calculated directly from independently measured system parameters without adjustable fitting coefficients. The experimentally observed dependence quantitatively follows the theoretical prediction, confirming that shorter memory times reduce temporal averaging and thereby enhance stochastic baseline variance. The corresponding timing jitter measurements are summarized in Fig. 5(b). The measured single-photon timing jitter exhibits the same dependence on the recovery time constant as the baseline fluctuations, increasing systematically as $\tau_e$ decreases. The correlated evolution of baseline variance and timing jitter further supports the theoretical prediction that the count-rate-dependent timing degradation originates from stochastic baseline dynamics.

### C. Influence of device dynamical properties

As shown in Fig. 6(a), devices with lower kinetic inductance exhibit faster electrical relaxation dynamics and correspondingly narrower output pulses. As a result, the time-integrated perturbation associated with each detection event is reduced, corresponding to a smaller effective perturbation parameter $Q$ in the stochastic theory.

The impact of this effect on stochastic baseline fluctuations is shown in Fig. 6(b). Devices with lower kinetic inductance exhibit systematically reduced baseline fluctuations across the entire count-rate range. This behavior confirms that stochastic baseline accumulation is governed by the strength of the perturbation introduced by each detection event.

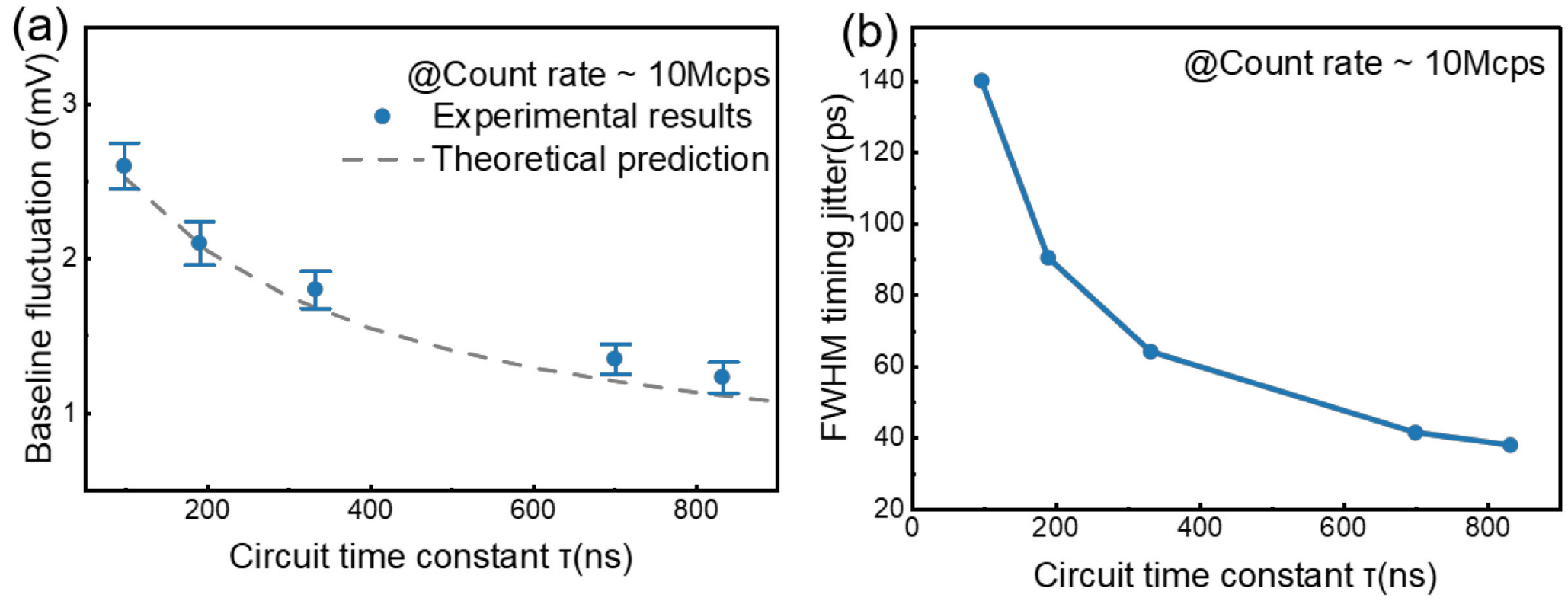

FIG. 5. Dependence of stochastic baseline fluctuations and timing jitter on the effective recovery time constant of the readout circuit. All measurements are performed at a fixed count rate of 10 Mcps. Error bars denote the standard deviation obtained from repeated measurements (N = 6). (a) Measured standard deviation of the pre-event baseline as a function of the effective recovery time constant $\tau_e$. The experimental results quantitatively follow the theoretical prediction without adjustable fitting parameters. Shorter recovery times produce larger baseline fluctuations due to reduced temporal averaging of residual responses. (b) Measured timing jitter FWHM of the single-photon peak as a function of $\tau_e$. The timing jitter follows the same dependence on recovery dynamics as the baseline fluctuations, demonstrating the direct conversion of stochastic baseline fluctuations into timing uncertainty.

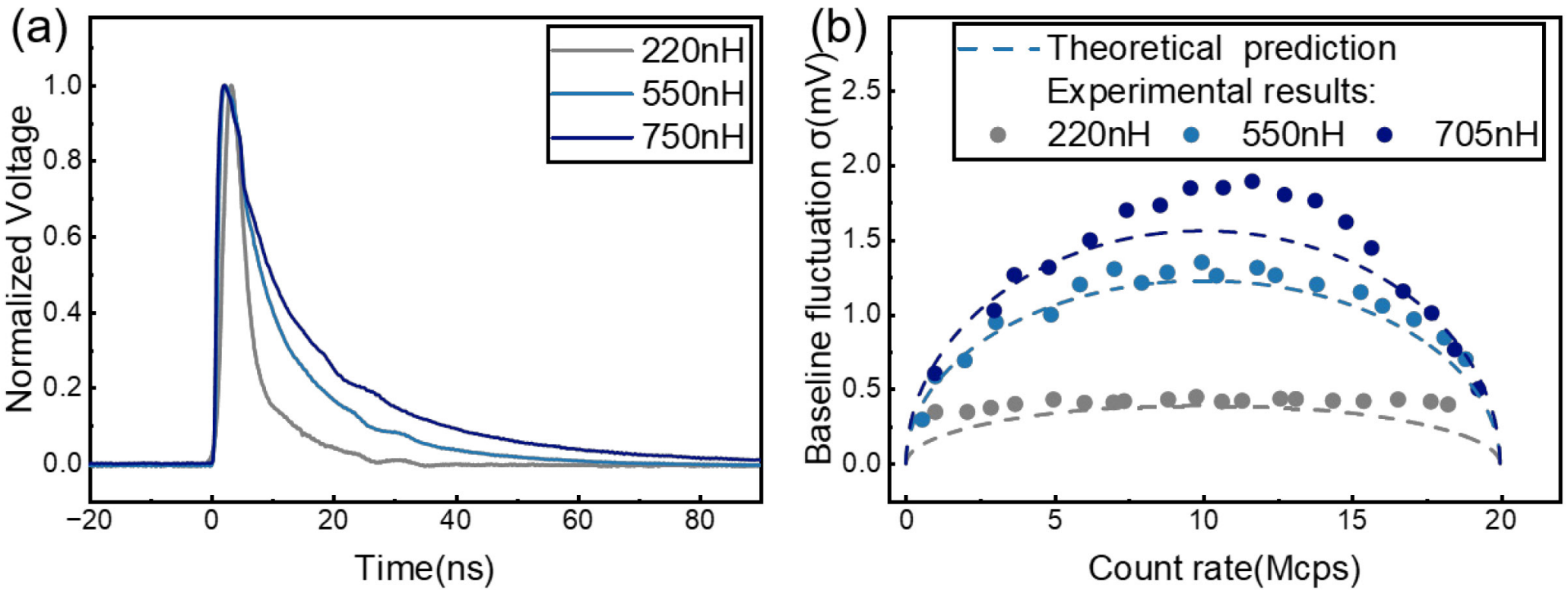

FIG. 6. Influence of detector dynamical properties on stochastic baseline fluctuations. (a) Normalized output pulse waveforms for SNSPD devices with different kinetic inductance values (220, 550, and 750 nH). Devices with lower kinetic inductance exhibit faster electrical relaxation and narrower pulse responses, corresponding to reduced time-integrated perturbation per detection event. (b) Measured baseline standard deviation as a function of count rate for devices with different kinetic inductance. Markers represent experimental data and dashed curves represent theoretical predictions. Devices with lower kinetic inductance exhibit systematically reduced baseline fluctuations due to weaker single-event perturbations. For the lowest-inductance device, the baseline fluctuations approach the electronic noise floor, resulting in weak count-rate dependence. For the highest-inductance device, systematic deviation from the theoretical prediction emerges at high count rates due to incomplete detector recovery and additional correlations between successive detection events.

For the device with the smallest kinetic inductance (220nH), the baseline standard deviation exhibits only a weak dependence on count rate. This indicates that the baseline fluctuation is masked by the system noise floor, placing the system in a noise-limited regime rather than a fluctuation-limited regime.

For the device with the largest kinetic inductance, systematic deviation from the theoretical prediction is observed at high count rates. This deviation originates from the breakdown of the independent-arrival approximation underlying the stochastic theory. In this regime, the detector recovery time constant becomes comparable to the laser repetition period $T = 50ns$, leading to incomplete detector recovery and additional correlations between successive detection events. The observed deviation therefore marks the crossover from a readout-dominated stochastic regime to a detector-dynamics-dominated regime[26, 27].

## V. DISSCUSSION

The stochastic framework developed in this work reveals a timing jitter mechanism originating from the interplay between stochastic excitation and finite-memory constraints imposed by the readout chain. In systems with a finite recovery time constant $\tau_e$, the baseline evolves into a stochastic dynamical variable driven by the accumulation of random detection events. Through threshold-based timing extraction, these baseline fluctuations are directly converted into timing uncertainty.

This mechanism is fundamentally distinct from previously identified effects such as multiphoton responses and pulse pile-up. Multiphoton effects are governed by photon-number statistics and primarily modify waveform shape, whereas pulse pile-up originates from deterministic temporal overlap between adjacent events. In contrast, the mechanism identified here arises from the stochastic accumulation of single-event responses in a finite-memory system, producing timing broadening even in the absence of strong multiphoton distortion or direct pulse overlap.

The present theory considers the experimentally relevant regime in which the readout response is dominated by a characteristic recovery time constant $\tau_e$. In realistic SNSPD readout systems, frequency-dependent impedance and multiple intrinsic relaxation processes may produce deviations from an ideal single-exponential response. Nevertheless, the experimentally observed scaling behavior is consistently reproduced across multiple independently controlled parameters, indicating that the experimentally observed timing broadening is governed primarily by stochastic finite-memory baseline dynamics.

The scaling behavior $\sigma_{t_{pulse}} \propto t_{rise} \cdot \tau_s / \sqrt{\tau_e} \cdot \sqrt{\lambda(1 - \lambda/f_{rep})}$ predicted by the stochastic framework further clarifies the physical parameters governing high-count-rate timing performance. Neglecting numerical prefactors and focusing on the dominant scaling relations, the baseline-induced timing broadening increases approximately with the square root of count rate and with the effective temporal perturbation associated with each detection event, while decreasing with the square root of the readout recovery timescale. Physically, larger single-event pulse areas introduce stronger perturbations into the finite-memory readout chain, whereas faster signal rise slopes reduce the sensitivity of threshold-crossing extraction to baseline variations. At the same time, longer recovery dynamics effectively average stochastic fluctuations over extended temporal windows, thereby suppressing baseline variance. These scaling relations provide a direct physical picture connecting detector pulse dynamics, readout memory effects, and timing performance in high-speed photon-counting systems.

From a system-design perspective, the present results highlight an inherent trade-off in high-speed readout architectures. AC coupling is widely employed to suppress low-frequency noise and stabilize broadband amplifier operation, but it simultaneously introduces an effective high-pass response with finite-memory recovery dynamics. Although DC coupling can partially relax this constraint, it often introduces long-term baseline drift and reduced operating stability. Consequently, practical high-speed SNSPD systems frequently adopt hybrid readout architectures. For example, previously reported high-count-rate SNSPD systems[23] employ DC coupling in the front-end stage while retaining AC-coupled amplification in subsequent stages, indicating that effective high-pass filtering remains difficult to eliminate entirely in realistic high-bandwidth implementations. The stochastic baseline fluctuations identified here should therefore be regarded as a generic feature of finite-memory high-speed photon-counting systems rather than an implementation-specific artifact.

## VI. CONCLUSIONS

In this work, we establish stochastic baseline fluctuations as an intrinsic mechanism of timing jitter broadening in high-count-rate SNSPD systems. Under stochastic photon arrival, finite-memory responses in the readout chain accumulate statistically and drive the baseline into a fluctuating dynamical state, which is directly converted into timing uncertainty through threshold-based discrimination.

We develop a unified stochastic framework that quantitatively connects photon-arrival statistics, readout dynamics, and timing performance under both continuous-wave and pulsed excitation. The theory predicts characteristic scaling laws, including the non-monotonic fluctuation behavior observed under pulsed excitation, and these predictions are quantitatively confirmed through systematic control of count rate, recovery time constant $\tau_e$, and detector dynamical properties.

These results establish stochastic baseline dynamics as a fundamental system-level contribution to timing broadening that is distinct from conventional multiphoton-response and pulse-pile-up mechanisms. More generally, the present work reveals that timing performance in high-speed photon-counting systems is fundamentally governed by the interplay between stochastic event statistics, finite system memory, and single-event response dynamics.

From a practical perspective, the present theory provides concrete design principles for high-speed, low-jitter photon-counting systems. In particular, suppressing low-frequency finite-memory responses, minimizing the time-integrated perturbation associated with individual detection events, and maximizing the signal slew rate near the discrimination threshold can systematically reduce baseline-induced timing broadening. These results provide a unified framework for the co-optimization of detector dynamics and readout electronics, offering a pathway toward next-generation high-throughput ultrafast photon-timing systems.

## ACKNOWLEDGMENTS

This work was supported by the Quantum Science and Technology-National Science and Technology Major Project (No. 2023ZD0300100), National Natural Science Foundation of China (U24A20320, 62401554).

## APPENDIX A. Experimental setup and SNSPD characterization

The experimental setup is illustrated in Fig. A1(a), including laser excitation, detector biasing, cryogenic amplification, and signal readout. Optical excitation is provided by a 1550 nm femtosecond pulsed laser source (Carmel X-1550, repetition frequency 20 MHz). The optical power is adjusted using fiber attenuators and a three-paddle polarization controller, calibrated by a power meter, before coupling into a self-aligned packaged SNSPD.

The detector is current-biased through the DC port of a bias tee, while the output pulses are amplified using a custom cryogenic amplifier and recorded either by a high-bandwidth oscilloscope (LeCroy WavePro 804HD) or a Time Tagger Ultra time-to-digital converter (TDC). The SNSPD is operated at a base temperature of approximately 0.85 K. The timing jitter of the femtosecond laser source is negligible compared with the measured system jitter. The TDC introduces an intrinsic timing uncertainty of approximately 10 ps FWHM, which contributes in quadrature to the measured timing resolution. Figure A1(b) shows a scanning electron micrograph of the fabricated SNSPD. The detector consists of a double-layer NbN nanowire[33, 34] integrated on a Si substrate with a high-reflectivity backside optical cavity[35]. The device has an active area of 12 μm, a nanowire width of 85 nm, and a pitch of 210 nm. The switching current is approximately 26.8 μA, and the intrinsic recovery time constant is approximately 11.4 ns.

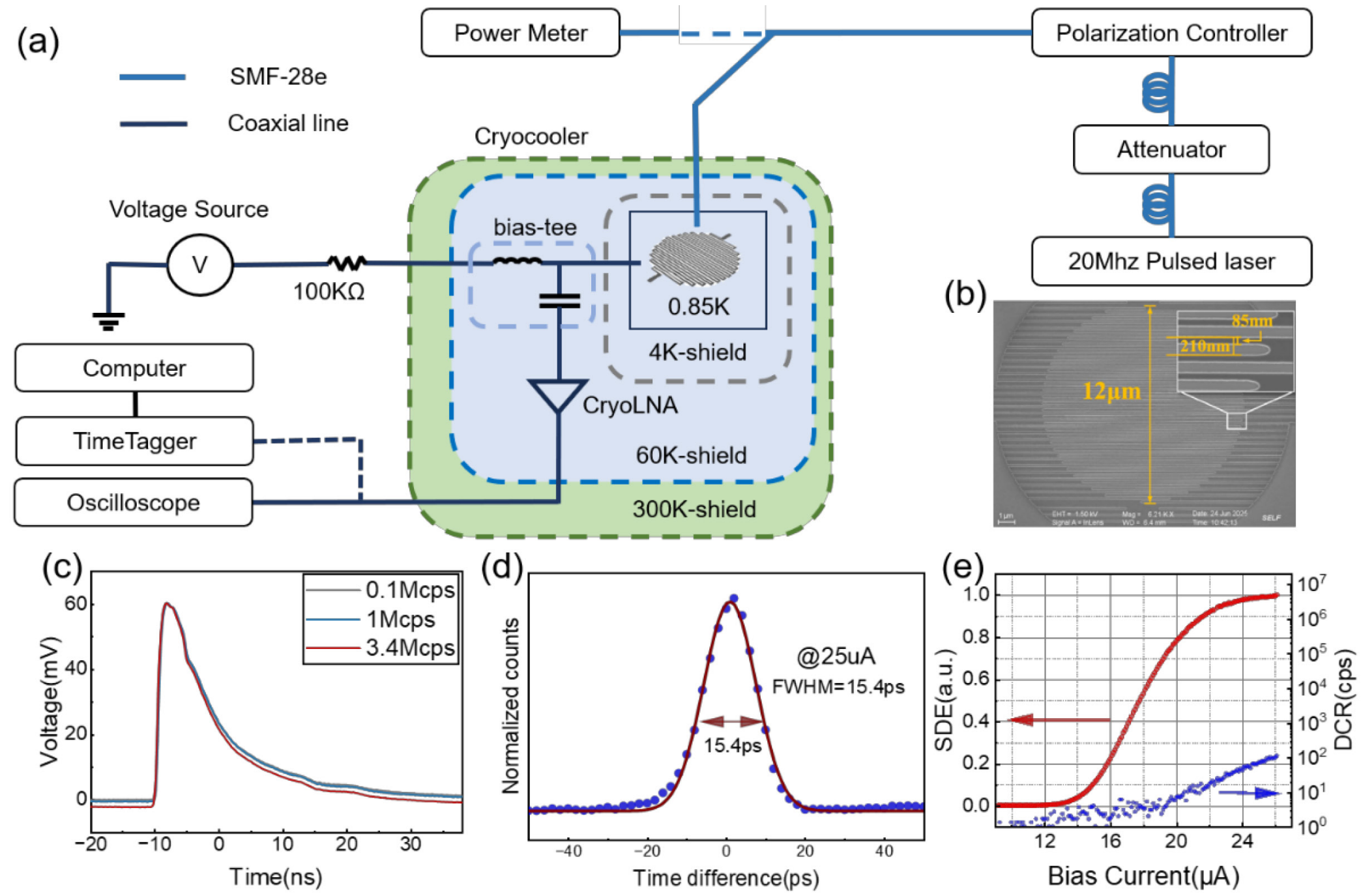

FIG. A1. (a) Experimental setup for SNSPD timing jitter characterization, including optical excitation, detector biasing, cryogenic amplification, and electronic readout. (b) Scanning electron micrograph of the fabricated SNSPD device. (c) Measured pulse waveforms at different count rates, showing progressively enhanced baseline fluctuations at high count rates due to finite-memory accumulation. (d) Measured timing distribution at a bias current of 25 μA, yielding a system timing jitter of 15.4 ps FWHM. (e) Normalized system detection efficiency and dark count rate as a function of bias current. The device exhibits a maximum system detection efficiency of approximately 75%.

Representative pulse waveforms measured at different count rates are shown in Fig. A1(c). As the count rate increases, progressively enhanced baseline fluctuations are observed due to the accumulation of finite-memory recovery responses. Figure A1(d) shows the measured timing distribution at a bias current of 25.0 μA and a count rate of 0.1 MHz, yielding a system timing jitter of 15.4 ps FWHM.

The normalized system detection efficiency and dark count rate as a function of bias current are presented in Fig. A1(e). The SNSPD used has an intrinsic detection efficiency close to saturation and a negligible dark count compared with the experimental photon count rate.

## APPENDIX B. Discrete-time stochastic theory under pulsed excitation

In this Appendix, we extend the stochastic finite-memory framework developed in the main text to the case of pulsed laser excitation. The system is naturally described in discrete time, where photon arrivals are synchronized with a repetition period $T = 1/f_{rep}$.

The impulse response of a single detection event is assumed to be unchanged from the continuous-time case and is given by

$$h(t) = \frac{Q}{\tau_e} \cdot e^{-\frac{t}{\tau_e}} \cdot u(t) \qquad (A1)$$

where $Q$ characterizes the time-integrated perturbation induced by a single event and $\tau_e$ is the recovery time constant of the readout chain.

Under pulsed excitation, photon arrivals are described by a discrete Bernoulli process. In each repetition period, a detection event occurs with probability

$$p = \frac{\lambda}{f_{rep}} \qquad (A2)$$

and the random variable $N_n$ describing the the $n$-th period takes values

$$N_n = \begin{cases} 1, & with\, probability\; p, \\ 0, & with\, probability\; 1-p. \end{cases} \qquad (A3)$$

The baseline is then described as the superposition of all past impulse responses,

$$B(t) = \sum_n N_n h(t - nT) \qquad (A4)$$

and the baseline immediately before each pulse is defined as

$$B_n = B(nT^-) \quad (A5)$$

From the finite-memory property of the impulse response, the baseline evolution follows the discrete recursion relation

$$B_{n+1} = (B_n + \frac{Q}{\tau_e} N_n) \cdot e^{-\frac{T}{\tau_e}} \quad (A6)$$

Defining $a = e^{-T/\tau_e}$, the mean baseline is obtained from the steady-state condition and takes the form:

$$\mu = a\mu + \frac{aQ}{\tau_e} p \overset{p=\frac{\lambda}{f_{rep}}}{\Longrightarrow} = \frac{Q}{\tau_e} \cdot \frac{\lambda}{f_{rep} \cdot \left(e^{\frac{T}{\tau_e}} - 1\right)} \quad (A7)$$

In the limit $T \ll \tau_e$, this expression reduces to

$$\mu = \lambda Q \quad (A8)$$

which recovers the continuous-time Poisson-driven result presented in the main text. To characterize fluctuations, we define the zero-mean variable

$$\tilde{B}_n = B_n - \mu \quad (A9)$$

Substituting into the recursion relation, we obtain

$$\tilde{B}_{n+1} = a\tilde{B}_n + \frac{aQ}{\tau_e}(N_n - p) \quad (A10)$$

The variance evolution follows

$$\sigma_{n+1}^2 = a^2\sigma_n^2 + \left(\frac{aQ}{\tau_e}\right)^2 Var(N_n) \quad (A11)$$

Using $Var(N_n) = p(1-p)$ , Under steady-state conditions $\sigma_{n+1}^2 = \sigma_n^2 = \sigma_B^2$, yielding

$$\sigma_B{}^2 = \frac{a^2Q^2p(1-p)}{\tau_e{}^2 \cdot (1-a^2)} \quad (A12)$$

Substituting $a = e^{-T/\tau_e}$ and $p = \lambda/f_{rep}$, the baseline standard deviation becomes

$$\sigma_B = \frac{Q}{\tau_e \cdot \sqrt{e^{\frac{2T}{\tau_e}} - 1}} \cdot \sqrt{\frac{\lambda}{f_{rep}}\left(1 - \frac{\lambda}{f_{rep}}\right)} \quad (A13)$$

In the limit $T \ll \tau_e$, we use the approximation

$$e^{\frac{2T}{\tau_e}} - 1 \approx \frac{2T}{\tau_e} \quad (A14)$$

the expression reduces to

$$\sigma_B \approx \frac{Q}{\sqrt{2\tau_e}} \cdot \sqrt{\lambda\left(1 - \frac{\lambda}{f_{rep}}\right)} \quad (A15)$$

Finally, in the limit $f_{rep} \to \infty$, the discrete excitation converges to the continuous Poisson limit, recovering

$$\sigma_B = Q\sqrt{\frac{\lambda}{2\tau_e}} \quad (A16)$$

which is consistent with the continuous-time result in the main text.

## APPENDIX C. Baseline definition and extraction procedure

The baseline used in this work is defined as the instantaneous signal level immediately preceding each detection event, consistent with the definition adopted in the stochastic theory developed in the main text.

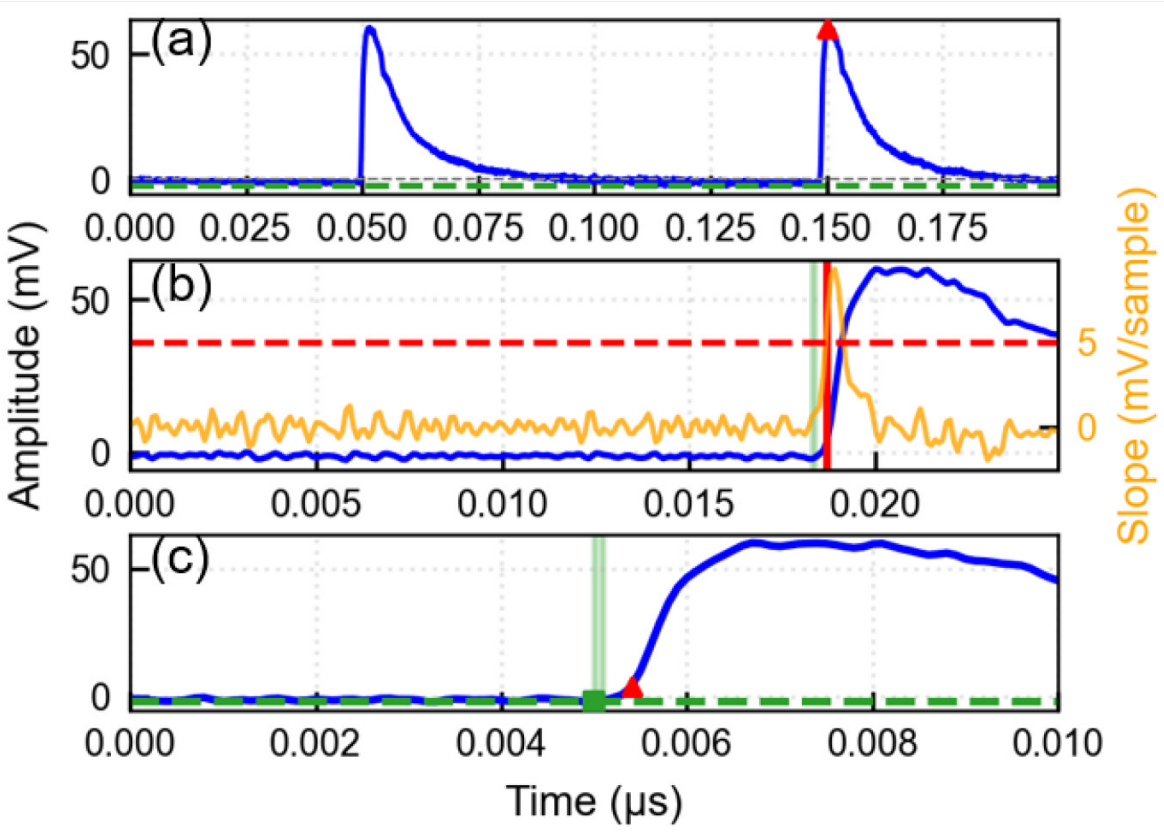

FIG. A2. Baseline definition and extraction procedure. (a) Pulse peak detection based on waveform maxima exceeding a 30 mV threshold. (b) Pulse onset determination using a slope criterion (first point exceeding 5 mV per sample). (c) Baseline assignment as the signal value 0.3–0.5 ns before the identified onset.

The experimental extraction procedure is illustrated in Fig. A2. Pulse peaks are first identified by locating waveform maxima exceeding a fixed discrimination threshold of 30 mV, as shown in Fig. A2(a). For each detected event, the pulse onset is determined using a slope-based criterion, defined as the first sample for which the rising slope exceeds 5 mV per sample, as shown in Fig. A2(b). The baseline value is then assigned as the signal amplitude at a fixed temporal offset of 0.3–0.5 ns prior to the onset point, as shown in Fig. A2(c).

This procedure ensures a consistent estimation of the pre-event baseline under high count-rate operation, where incomplete recovery of preceding pulses may otherwise bias baseline estimation if a broader temporal window is used. The resulting baseline definition is directly compatible with the theoretical framework, enabling quantitative comparison between measured fluctuations and theoretical predictions.

## APPENDIX D. Extraction of theoretical prediction parameters

The cryogenic amplifier provides a broadband gain of approximately 36 dB, corresponding to a voltage gain of ~63. Combined with the 50 Ω readout impedance $R_{load}$ , the peak pulse amplitude is approximately

$$A \approx I_{bias} \cdot 50\Omega \cdot 63 \quad (A17)$$

The characteristic pulse decay time $\tau_s$ is extracted directly from the measured waveform as the $1/e$ relaxation time. The effective perturbation parameter

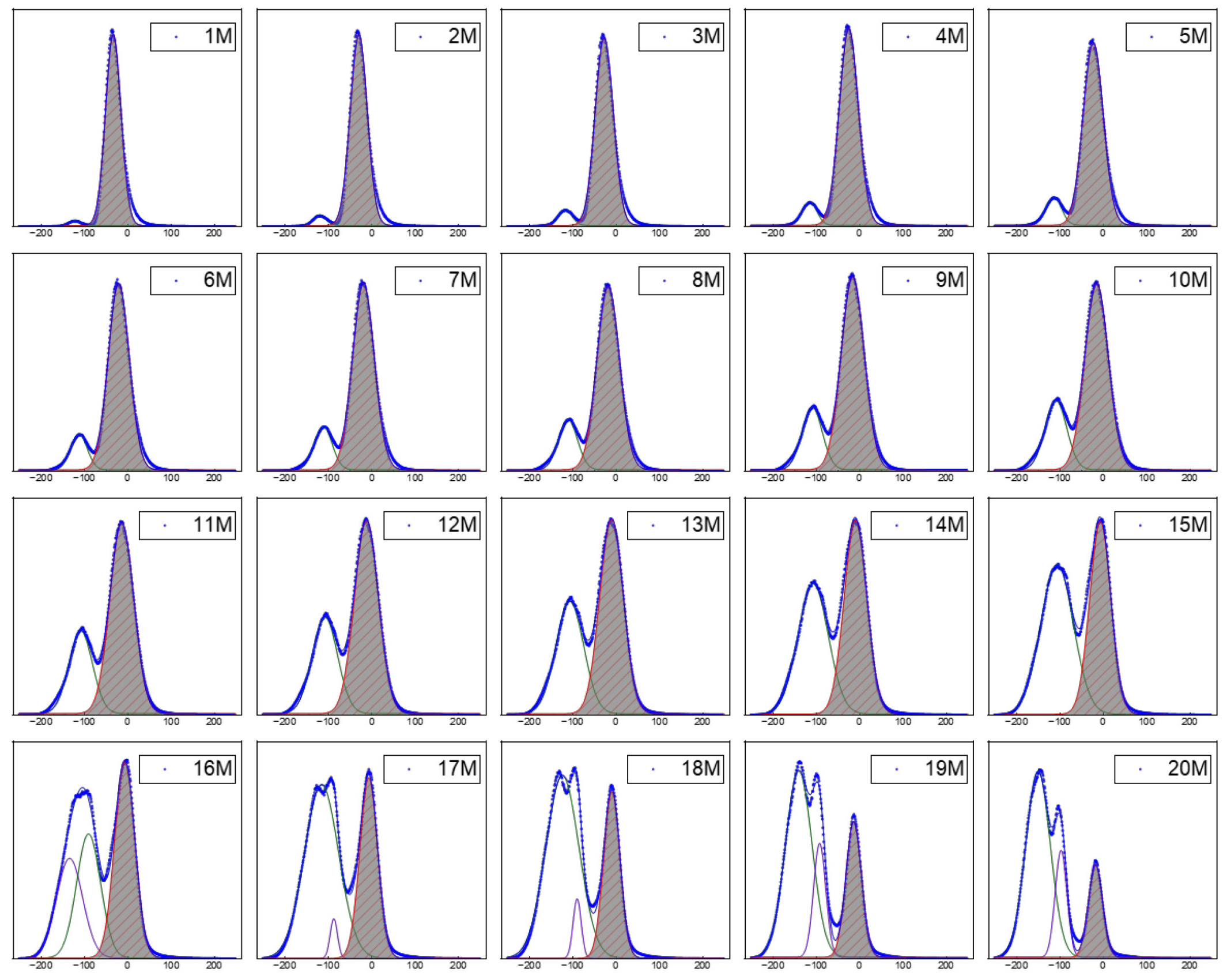

FIG. A3. Timing histogram under high-count-rate operation. Multiple peaks arising from multiphoton detection events are observed at high count rates[26, 28]. The rightmost peak corresponds to the single-photon response and is used for timing jitter extraction.

entering the stochastic theory is then estimated from the time-integrated pulse response,

$$Q \approx A\tau_s \tag{A18}$$

which characterizes the cumulative baseline perturbation induced by a single detection event.

Throughout the experiments, the SNSPD readout exhibits finite-memory recovery dynamics governed by the AC-coupled response of the readout chain. The amplifier possesses an intrinsic capacitance $C_{amp}$ of approximately 20 nF, which combines in series with the external coupling capacitor $C_{ext}$ to produce an effective capacitance

$$C_{eff} = \frac{C_{ext} \cdot C_{amp}}{C_{ext} + C_{amp}} \tag{A19}$$

and corresponding recovery time constant

$$\tau_e = R_{load} \cdot C_{eff} \tag{A20}$$

This RC response determines the dominant low-frequency recovery dynamics governing the stochastic baseline fluctuations throughout the measurements.

## APPENDIX E. Extraction of single-photon timing jitter

At high count rates, the timing histogram exhibits multiple peaks due to multiphoton detection events[26], as shown in Fig. A3. These peaks correspond to different photon-number responses and result in an overall broadening of the timing distribution. The rightmost peak is attributed to the single-photon response and is used for jitter extraction, as it corresponds to the earliest detection events in the multiphoton hierarchy.

To isolate the single-photon contribution, the timing histogram is fitted using a multi-Gaussian model. Each peak is represented by a Gaussian function, and the overall distribution is obtained through least-squares minimization. The fitting stability is verified by the insensitivity of extracted peak parameters to variations in fitting range.

The timing jitter is defined as the full width at half maximum (FWHM) of the rightmost Gaussian component. This definition removes multiphoton-induced broadening and provides a consistent metric for comparison with theoretical predictions.

## APPENDIX F. Validation under different excitation regimes

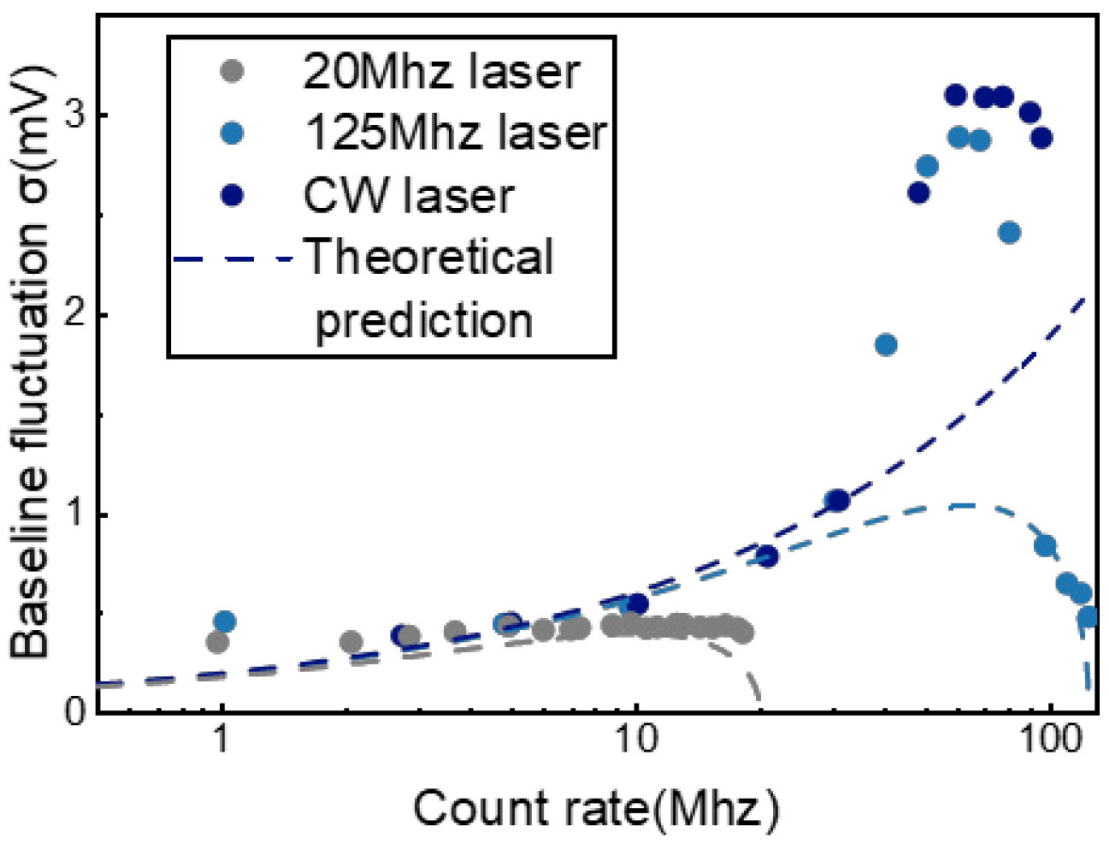

FIG. A4. Comparison of baseline fluctuation scaling under different excitation conditions. Measured baseline standard deviation as a function of count rate under 20 MHz pulsed excitation, 125 MHz pulsed excitation, and continuous-wave (CW) operation. Experimental data are compared with theoretical predictions from the stochastic-process framework.

Figure A4 shows the measured baseline standard deviation as a function of count rate under 20 MHz pulsed excitation, 125 MHz pulsed excitation, and continuous-wave (CW) operation. For the 125 MHz measurements, an SNSPD device with a faster intrinsic recovery time constant (~4.4 ns) is used to reduce pulse overlap effects associated with the shorter repetition period. The detector bias current is maintained at 10 μA for all measurements.

As shown in Fig. A4, the experimental results under all excitation conditions follow the characteristic count-rate dependence predicted by the stochastic framework developed in the main text. For the 125 MHz and CW regimes, the measured baseline fluctuations deviate upward from the ideal prediction at high count rates.

This deviation arises from the breakdown of the assumption of independent detection events in the stochastic framework. In this regime, the inter-event interval becomes comparable to the detector recovery time constant, leading to incomplete relaxation of the detector between successive events. Consequently, detector recovery dynamics introduce additional correlations that enhance baseline fluctuations beyond the readout-limited prediction.

Overall, these results confirm that the stochastic accumulation mechanism remains valid across different excitation regimes, while identifying the regime in which detector dynamics become non-negligible.